# Exact solvability of two new 3D and 1D nonrelativistic potentials within the TRA framework

I. A. Assi[a,1], H. Bahlouli[b,c], A. Hamdan[d]

[a] *Department of Physics and Physical Oceanography, Memorial University of Newfoundland, St. John's, NL, A1B 3X7, Canada*
[b] *Physics Department, King Fahd University of Petroleum & Minerals, Dhahran 31261, Saudi Arabia*
[c] *Saudi Center for Theoretical Physics, P.O. Box 32741, Jeddah 21438, Saudi Arabia*
[d] *Department of Mathematics and Statistics, Memorial University of Newfoundland, St. John's, NL, A1C 5S7, Canada*

**Abstract**: This work is concerned about introducing two new 1D and 3D confined potentials and present their solutions using the Tridiagonal Representation Approach (TRA). The wavefunction is written as a series in terms of square integrable basis functions which are expressed in terms of Jacobi polynomials. Moreover, the expansion coefficients are written in terms of new orthogonal polynomials that were introduced recently by Alhaidari, the analytical properties of these polynomials are yet to be derived. Moreover, we have computed the numerical eigen-energies for both potentials by considering specific choices of the potential parameters.

## 1. Introduction

Exactly solvable potentials in non-relativistic quantum mechanics enable us to gain much insight and deep understanding of physical phenomena due to the existence of a closed form solution of the quantum mechanical problem at hand. This is the reason why the search for exactly solvable potentials has been a subject of great importance since the early days of quantum mechanics [1]. Various methods have been devised for solving the stationary Schrodinger equation, the factorization method was originally introduced by Schrodinger [2] and is summarized in the classical work by Infeld and Hull [3]. The supersymmetric quantum mechanics approach (SUSY) generates exactly solvable quantum mechanical potentials in one dimension [4,6] and higher dimensions [7] using symmetry ideas inspired from field theory [5]. A comprehensive review on supersymmetric quantum mechanics can be found in the book [8]. The tridiagonal representation approach (TRA) is based on a suitable choice of a complete basis set that is compatible with the physical domain of Schrodinger equation and enable this equation to be represented by a three-term recursion relation for the wave function expansion coefficient [9], hence building a strong link with the rich domain of orthogonal polynomials in mathematics literature.

In this work, we are interested in extending the solution space of Schrodinger equation by solving the wave equation for the following 3D and 1D confined potentials using the TRA

---

[1] Corresponding Author: Email Address: iassi@mun.ca

$$V(r) = \frac{V_0}{\sinh^2(\lambda r/2)} + \frac{V_1}{\cosh^2(\lambda r/2)} + V_2 \cosh(\lambda r), \tag{1.1}$$

where $0 \leq r < \infty$, $V_0, V_2 > 0$ and $V_1 < 0$ for bound states as discussed in section 2, see Eq. (2.10) and below, and

$$V(x) = \frac{V_1}{\cosh^2(\lambda x/2)} + V_2 \cosh(\lambda x), \tag{1.2}$$

with $-\infty < x < +\infty$. For the potentials in (1.1) and (1.2), when considering $V_2 = 0$, it reduces to the well-known hyperbolic Pöschl-Teller potential which was solved analytically in the literature using different methods [10, 11]. Thus, in our present work, we will only consider the case when $V_2 \neq 0$, which, up to our knowledge, was never reported in literature.

The organization of this article goes as follows. In section 2, we present the TRA formulation of the problem with the focus on the potential (1.1). Then, in the third section, we give an approximate analytical solution of Schrodinger equation for nonzero angular momentum case for the 3D confined potential. On the other hand, the potential (1.2) is discussed in detail in the fourth section. Finally, we present our conclusion in the fifth section.

## 2. TRA Derivations

Starting with the effective stationary 1D radial Schrodinger equation [12-14]

$$\left\{ -\frac{1}{2}\frac{d^2}{dr^2} + V(r) - E \right\} \psi(r) = 0, \tag{2.1}$$

where $0 \leq r < \infty$, $V(r)$ is the potential energy. Using the coordinate transformation $y = \cosh(\lambda r)$, we transform Eq. (2.1) to the following form

$$\left\{ \frac{\lambda^2(1-y^2)}{2}\frac{d^2}{dy^2} - \frac{\lambda^2 y}{2}\frac{d}{dy} + V(y) - E \right\} \psi(y) = 0, \tag{2.2}$$

Next, we expand the wavefunction in terms of square integrable basis functions $\{\chi_k(y)\}_{k=0}^{\infty}$ as $\psi_n(y) = \sum_k f_k(E) \chi_k(y)$, where $\{f_k(E)\}_{k=0}^{\infty}$ are functions of the energy and the potential parameters which to be determined later [15-18]. Here we are assuming a discrete energy spectrum, in case of the presence of an additional continuous spectrum the wave function will be expanded in both discrete and continuous Fourier components keeping in mind that discrete and continuous energy spectra do not overlap. A suitable choice of basis functions for $y \geq 1$ is given in terms of Jacobi polynomials as follows [19]

$$\chi_k(y) = A_k (y-1)^\alpha (y+1)^\beta P_k^{(\mu,\nu)}(y), \tag{2.3}$$

where

$A_k = \sqrt{(2k+\mu+\nu+1)\Gamma(k+1)\Gamma(k+\mu+\nu+1)\sin\pi(\mu+\nu+1)/2^{\mu+\nu+1}\Gamma(k+\nu+1)\Gamma(k+\mu+1)\sin\pi\nu}$

is a normalization constant (see Eq. (2.4) below), $\{\alpha, \beta, \mu, \nu\}$ are real potential parameters with $\mu > -1$ and $\mu + \nu < -2N - 1$ for some integer $N \gg 1$. The square integrability is guaranteed by the following orthogonality relation [19]

$$\int_1^\infty (y-1)^\mu (y+1)^\nu P_k^{(\mu,\nu)}(y) P_l^{(\mu,\nu)}(y) dy = A_k^{-2} \delta_{kl}, \tag{2.4}$$

Defining the matrix representation of the wave operator in (2.1) as

$$\Im_{kl} = \int_0^\infty dr\, \chi_k(r) \left[ -\frac{1}{2}\frac{d^2}{dr^2} + V(r) - E \right] \chi_l(r)$$

$$= \frac{\lambda}{2} \int_1^\infty dy (y^2-1)^{-\frac{1}{2}} \chi_k(y) \left[ (1-y^2)\frac{d^2}{dy^2} - y\frac{d}{dy} + U(y) - \varepsilon \right] \chi_l(y) \tag{2.5}$$

where $U(r) = 2V(r)/\lambda^2$ and $\varepsilon = 2E/\lambda^2$. Now, using (2.3) in (2.5), with few algebraic steps, we write the matrix representation (2.5) as

$$\Im_{kl} = \frac{\lambda A_k A_l}{2} \int_1^\infty dy (y-1)^{2\alpha-1/2} (y+1)^{2\beta-1/2} P_k^{(\mu,\nu)}(y) \left[ (1-y^2)\frac{d^2}{dy^2} - \left[(2\alpha + 2\beta + 1)y + 2\alpha - 2\beta\right]\frac{d}{dy} \right] P_l^{(\mu,\nu)}(y)$$

$$+ \frac{\lambda A_k A_l}{2} \int_1^\infty dy (y-1)^{2\alpha-1/2} (y+1)^{2\beta-1/2} P_k^{(\mu,\nu)}(y) \left[ \frac{\alpha(2\alpha-1)}{1-y} + \frac{\beta(2\beta-1)}{1+y} - (\alpha+\beta)^2 + U(y) - \varepsilon_n \right] P_l^{(\mu,\nu)}(y) \tag{2.6}$$

Using the orthogonality relation (2.4), along with the differential equation for the Jacobi polynomials we the require the above matrix element to be tridiagonal, this will result in three possibilities:

a) $\mu = 2\alpha - 1/2$, and $\nu = 2\beta - 1/2$, (2.7a)
b) $\mu = 2\alpha - 1/2$, and $\nu = 2\beta - 3/2$, (2.7b)
c) $\mu = 2\alpha - 3/2$, and $\nu = 2\beta - 1/2$, (2.7c)

The potential functions in the last two cases are singular at the origin with an inverse square singularity. In fact, they correspond to the well-known class of exactly solvable potentials, the hyperbolic Pöschl-Teller potential, $V(r) = \frac{V_1}{\cosh^2(\lambda r)} + \frac{V_2}{\sinh^2(\lambda r)}$. Consequently, we ignore those two cases and focus on the first case which results in

$$\Im_{kl} = -\frac{\lambda}{2}\left[\varepsilon_n + k(k+\mu+\nu+1) + \left(\frac{\mu+\nu+1}{2}\right)^2\right]\delta_{kl} + \frac{\lambda}{2} A_k A_l \times$$

$$\int_1^\infty dy (y-1)^\mu (y+1)^\nu P_k^{(\mu,\nu)}(y) \left[\frac{1}{2}\frac{\mu^2-1/4}{1-y} + \frac{1}{2}\frac{\nu^2-1/4}{1+y} + U(y)\right] P_l^{(\mu,\nu)}(y) \tag{2.8}$$

Using the three-term recursion relation of Jacobi polynomials [20], we find that Eq. (2.8) is tridiagonal only for the following form of the potential function

$$U(y) = \frac{2U_-}{1-y} + \frac{2U_+}{1+y} + U_0 y + U_1, \tag{2.9}$$

where $U_- = -\frac{1}{4}(\mu^2 - 1/4)$, $U_+ = -\frac{1}{4}(\nu^2 - 1/4)$, $U_0$ and $U_1$ are real potential parameters. From now on, we set $U_1 = 0$. In the $r$ configuration, the potential (2.9) reads

$$V(r) = \frac{V_0}{\sinh^2(\lambda r/2)} + \frac{V_1}{\cosh^2(\lambda r/2)} + V_2 \cosh(\lambda r), \qquad (2.10)$$

where $V_0 = \lambda^2(\mu^2 - 1/4)/8$, $V_1 = -\lambda^2(\nu^2 - 1/4)/8$, and $V_2 = \lambda^2 U_0/2$. Consequently, the matrix representation of the wave operator is

$$\mathfrak{I}_{kl} = -\frac{\lambda}{2}\left[\varepsilon + \left(k + \frac{\mu+\nu+1}{2}\right)^2\right]\delta_{kl} + \frac{\lambda}{2}U_0 \langle k|y|l\rangle, \qquad (2.11)$$

where $\langle k|y|l\rangle = A_k A_l \times \int_1^\infty dy\,(y-1)^\mu (y+1)^\nu P_k^{(\mu,\nu)}(y) y P_l^{(\mu,\nu)}(y)$. Clearly, if $U_0 = 0$, the matrix in (2.11) becomes diagonal and the energy spectrum is given below

$$\varepsilon_n = -\left(n + \frac{\mu+\nu+1}{2}\right)^2, \qquad (2.12)$$

Using $V_0 = \lambda^2(\mu^2 - 1/4)/8$ and $V_1 = -\lambda^2(\nu^2 - 1/4)/8$, the above equation becomes

$$\varepsilon_n = -\frac{1}{4}\left(2n + 1 + \sqrt{\frac{8V_0}{\lambda^2} + \frac{1}{4}} - \sqrt{\frac{1}{4} - \frac{8V_1}{\lambda^2}}\right)^2, \qquad (2.13)$$

which is the same spectrum formula of the hyperbolic Pöschl-Teller potential [10, 11]. Knowing that the hyperbolic Pöschl-Teller potential has a continuous energy spectrum or a mix of discrete and continuous spectra depending on the values the potential parameters then n can be continuous or discrete in the above formula. In addition, the finite size of the spectrum (2.13) is explained as follows. Recall that in this formalism, Jacobi polynomials are defined as $P_n^{(\mu,\nu)}(y)$, where $n \in \{0,1,\ldots,N\}$. Now, using $\mu+\nu < -2N-1$ and the values of the parameters below (2.10), one can easily show that $N < \frac{1}{2}\left|\sqrt{\frac{1}{4} - \frac{8V_1}{\lambda^2}} - \sqrt{\frac{1}{4} + \frac{8V_0}{\lambda^2}} - 1\right|$ giving a finite maximum value for $N$ and hence a finite number of bound states. Proceeding with the case $U_0 \neq 0$, we use the three-term recursion relation of Jacobi polynomials together with the orthogonality relation (2.4), we get [20]

$$\langle k|y|l\rangle = -C_k \delta_{k,l} + D_{k-1}\delta_{k,l+1} + D_k \delta_{k,l-1}, \qquad (2.14)$$

where $C_k = \frac{\mu^2 - \nu^2}{(2k+\mu+\nu)(2k+\mu+\nu+2)}$ and $D_k = \frac{2}{2k+\mu+\nu+2}\sqrt{\frac{(k+1)(k+\mu+1)(k+\nu+1)(k+\mu+\nu+1)}{(2k+\mu+\nu+1)(2k+\mu+\nu+3)}}$. Using this result together with (2.11), we write the following three-term recursion relation of the coefficients $\{f_k(E)\}_{k=0}^\infty$

$$\left[\varepsilon_n + \left(k + \frac{\mu+\nu+1}{2}\right)^2 + U_0 C_k\right] f_k(\varepsilon) = U_0 \left[D_{k-1} f_{k-1}(\varepsilon) + D_k f_{k+1}(\varepsilon)\right], \tag{2.15}$$

By making the transformation $P_k(\varepsilon) = \frac{A_0}{A_k} \frac{f_k(\varepsilon)}{f_0(\varepsilon)}$ in (2.15), we get

$$\frac{1}{U_0}\left[\varepsilon + \left(k + \frac{\mu+\nu+1}{2}\right)^2\right] P_k(\varepsilon) = \frac{2(k+\mu)(k+\nu)}{(2k+\mu+\nu)(2k+\mu+\nu+1)} P_{k-1}(\varepsilon) + \frac{2(k+1)(k+\mu+\nu+1)}{(2k+\mu+\nu+1)(2k+\mu+\nu+2)} P_{k+1}(\varepsilon) - C_k P_k(\varepsilon), \tag{2.16}$$

Comparing the above equation with Eq. (9) in [21], we find that the general solution of (2.16) is given below

$$P_k(\varepsilon) = \bar{H}_k^{(\mu,\nu)}\left(-U_0^{-1}; \varepsilon, \frac{\pi}{2}\right), \tag{2.17}$$

where $\bar{H}_k^{(\mu,\nu)}(z^{-1}; \alpha, \theta)$ is a new polynomial in $z^{-1}$ and $\alpha$ of degree $k$ as introduced by Alhaidari in [21]. The analytical properties of this polynomials are still to be determined. However, we can obtain iteratively those polynomials up to any order using the initial conditions $\bar{H}_{-1}^{(\mu,\nu)}(z^{-1}; \alpha, \theta) = 0$ and $\bar{H}_0^{(\mu,\nu)}(z^{-1}; \alpha, \theta) = 1$. The general solution of Schrodinger equation (2.1) for the potential in (2.9) reads

$$\psi_\varepsilon(r) = C_n (y-1)^{\frac{2\mu+1}{4}} (y+1)^{\frac{2\nu+1}{4}} \sum_{k=0}^{N} \tilde{A}_k \bar{H}_k^{(\mu,\nu)}\left(-U_0^{-1}; \varepsilon, \frac{\pi}{2}\right) P_k^{(\mu,\nu)}(y), \tag{2.18}$$

where $y = \cosh(\lambda r)$, $C_n = C_n(\mu, \nu)$ is a normalization constant, and

$$\tilde{A}_k = \frac{(2k+\mu+\nu+1)\Gamma(k+1)\Gamma(k+\mu+\nu+1)}{\Gamma(k+\nu+1)\Gamma(k+\mu+1)}, \tag{2.19}$$

Clearly, the wavefunction in (2.18) satisfies the boundary conditions at $y = \{1, +\infty\}$ by requiring $\mu > -1/2$ (which is okay since $\mu > -1$), and $\nu < -1/2$ which is already satisfied since $\mu + \nu < -2N - 1$ for large positive integer $N$ [19].

In the next section we present an approximate solution of the wave equation for the case of nonzero angular momentum quantum number.

## 3. Approximate analytical solution of the wave equation for the arbitrary *l*-wave case

The radial Schrodinger equation reads

$$\left\{-\frac{1}{2}\frac{d^2}{dr^2} + V(r) + \frac{\ell(\ell+1)}{2r^2} - E\right\}\psi(r) = 0, \tag{3.1}$$

where we plotted the effective potential for different values of the angular momentum as shown in **Figure 1**. In the presence of centrifugal term, one needs to use approximations of the centrifugal term to allow analytical solutions of the wave equation (3.1). For the potential in (2.10), we use the following approximation scheme

$$\frac{1}{r^2} \approx \frac{\lambda^2}{4}\left[\frac{1}{\sinh^2(\lambda r/2)} + \frac{31}{945}\frac{1}{\cosh^2(\lambda r/2)} - \frac{16}{945}\cosh(\lambda r) + \frac{20}{63}\right], \quad (3.2)$$

where Taylor series of the RHS of (3.2) is $RHS = \frac{1}{r^2} + O(r^6)$. Consequently, the effective Schrodinger equation becomes

$$\left\{-\frac{1}{2}\frac{d^2}{dr^2} + \frac{\tilde{V}_0}{\sinh^2(\lambda r/2)} + \frac{\tilde{V}_1}{\cosh^2(\lambda r/2)} + \tilde{V}_2\cosh(\lambda r) - \tilde{E}\right\}\psi(r) = 0, \quad (3.3)$$

where $\tilde{V}_0 = V_0 + \frac{\lambda^2 \ell(\ell+1)}{8}$, $\tilde{V}_1 = V_1 + \frac{31}{945}\frac{\lambda^2 \ell(\ell+1)}{8}$, $\tilde{V}_2 = V_2 - \frac{16}{945}\frac{\lambda^2 \ell(\ell+1)}{8}$, and $\tilde{E} = E - \frac{20}{63}\frac{\lambda^2 \ell(\ell+1)}{8}$. The solution of (3.3) is simply as (2.18) but with $V_0 \to V_0 + \frac{\lambda^2 \ell(\ell+1)}{8}$, $V_1 \to V_1 + \frac{31}{945}\frac{\lambda^2 \ell(\ell+1)}{8}$, $V_2 \to V_2 - \frac{16}{945}\frac{\lambda^2 \ell(\ell+1)}{8}$ and $E \to E - \frac{20}{63}\frac{\lambda^2 \ell(\ell+1)}{8}$. Note that in this formalism $V_0 > 0$ since $\mu > -1/2$ which means the potential supports only bound states if we only consider $V_2 > 0$. However, for $V_2 < 0$ this potential is repulsive for large distances and does not support bound and scattering states. In **Figure 2c** where $V_1 > 0$ and $V_2 < 0$ the potential seems to support resonance states mathematically however due to the large potential barrier it will not be able to be realized practically. A more careful mathematical analysis however will be desirable to support these arguments.

Here, we are interested in $V_2 > 0$ in which the system has only bound states. The lack of analytic properties of the new polynomial $\bar{H}_n^{(\mu,\nu)}(z^{-1};\alpha,\theta)$ makes it hard to express the eigenvalues and the other properties of this system in compact form [21, 23]. Thus, we rely on numerical solutions at this stage. Generally, within the TRA, we usually rewrite the three-term recursion relation in matrix form as

$$T|f_k(\varepsilon)\rangle = \varepsilon|f_k(\varepsilon)\rangle, \quad (3.4)$$

where $T$ is a tridiagonal symmetric matrix whose elements for the current problem takes the following form

$$T_{kl} = -\left\{\left[\left(k + \frac{\mu+\nu+1}{2}\right)^2 - \frac{\ell(\ell+1)}{12}\right] + U_0 C_k\right\}\delta_{k,l} + U_0\left[D_{k-1}\delta_{k,l+1} + D_k\delta_{k,l-1}\right], \quad (3.5)$$

where $U_0 = 2V_2/\lambda^2$ and $\varepsilon = 2E/\lambda^2$. We vary the size of the matrix (3.5) until we observe convergence in its eigenvalues which are related to the actual energies of (3.3) by considering the eigenvalue equation (3.4). However, in this special case, the matrix size is constrained by the requirement $N < -\frac{\mu+\nu+1}{2}$ as the parameters $\{\mu,\nu\}$ are fixed by the choice of potential parameters. Consequently, we rely on other techniques to obtain the eigenenergies as in **Appendix**

**A**. As an illustration, we tabulated the lowest few energies for different choices of the potential parameters and different values of the angular momentum as shown in **Table 1** and **Table 2**. We should note that it is still possible to use the usual TRA method (Equations (3.4) and (3.5)) for certain choices of the potential parameters at which the maximum possible value of $N$ becomes large, see **Table 5** which is related to the potential (1.2) which will be discussed in the following section.

## 4. The infinite 1D hyperbolic potential well

It turns out that all the calculations in section 2 are correct even if the physical space becomes the whole real line because $1 \leq \cosh(\lambda x) < +\infty$ regardless of whether $x \in [0, +\infty)$ or $x \in (-\infty, +\infty)$. However, a factor of 2 appears in the $J$-matrix for this case which does not affect the overall calculations, i.e. $J_{n,m}(1D) = 2J_{n,m}(3D)|_{V_0=0}$. In this case, we are interested in solving the following 1D Schrodinger equation

$$\left\{ -\frac{1}{2}\frac{d^2}{dx^2} + \frac{V_1}{\cosh^2(\lambda x/2)} + V_2 \cosh(\lambda x) - E \right\} \psi(x) = 0, \tag{4.1}$$

which corresponds to the case $\mu^2 = 1/4$. As an illustration, we plot the potential (1.2) for different choices of the potential parameters as in **Figure 3**. It's clear that this potential is even in $x$ along the domain $x \in (-\infty, +\infty)$, which means that the corresponding states are even and odd in $x$ [13, 14]. The even states are associated with $\mu = -1/2$ and the odd states correspond to $\mu = +1/2$. We should point here that the basis functions are even in $x$ (see Eq. 2.3), but the physical requirement forces us to write the normalized eigen solutions of Eq. (4.1) as follows

$$\chi(x) = \frac{1}{\sqrt{2}}[\psi(x) \pm \psi(-x)], \tag{4.2}$$

where the plus sign corresponds to the even states, the minus is for the odd states, and $\psi(x)$ is the normalized solution of Eq. (4.1) which reads

$$\psi(x) = \frac{1}{A_0} f_0(\varepsilon)(y-1)^{\frac{2\mu+1}{4}}(y+1)^{\frac{2\nu+1}{4}} \sum_{k=0}^{N} \tilde{A}_k \bar{H}_k^{(\mu,\nu)}\left(-U_0^{-1}; \varepsilon, \frac{\pi}{2}\right) P_k^{(\mu,\nu)}(y), \tag{4.3}$$

with $y = \cosh(\lambda x)$. Recall that $\mu + \nu < -2N - 1$ implies $\nu < -2N - 1/2$ or $\nu < -2N - 3/2$, this guarantees that the wavefunction vanishes at the boundaries, i.e. $\psi(x = \pm\infty) = 0$. The calculation of the energy spectrum was done numerically using the same approach used for the 3D problem as explained in **Appendix A**. For the eigenvalues of (4.1), we have computed the lowest few energies for the following choice of parameters $V_1 = -100$, $V_2 = 5$, and $\lambda = \sqrt{2}$ as shown in **Tables 3** and **4** using the method of **Appendix A**. Also, we have calculated the lowest ten energies for $V_1 = -10000$, $V_2 = 5$, and $\lambda = \sqrt{2}$ using equations (3.4) and (3.5) with $\mu = -1/2$ and $\ell = 0$ as given in **Table 5**. In addition, we plotted the unnormalized wavefunction for the lowest four states as in **Figure 4** for the same parameters taken in **Tables 3** and **4**.

## 5. Conclusions

In this work, we have extended the class of solutions of Schrodinger equation with two new potentials using the TRA. The wavefunctions were expressed as a series in terms of Jacobi polynomials with the expansion coefficients written in terms of new orthogonal polynomials introduced by Alhaidari very recently [21]. The analytical properties of these polynomials are not yet known, however getting their properties in compact form allows us to write the corresponding bound state energy, scattering phase shift and other properties of the system in closed forms. We have also confirmed that the new potential (1.1) with $V_2 < 0$ will not support any bound or scattering states, even resonance state will not be able to be realized practically. A more careful mathematical analysis however will be desirable to support these arguments.

Finally, we look forward that our present findings can be used in some applications in real physical systems.

**Acknowledgments**

The authors would like to thank the referee of the valuable comments and suggestions that helped us in improving the manuscript. HB acknowledge the support of the Saudi Center for Theoretical Physics (SCTP).

**Appendix A: The Hamiltonian Diagonalization Method for obtaining the energy spectrums of the 1D and 3D potentials**

As discussed in the main sections, the absence of the analytic properties of the new energy polynomial together with the constraint $\mu + \nu < -2N - 1$ make it hard for us to follow the usual numerical methods used in the TRA for obtaining the eigenenergies. In this work, we relied on the Hamiltonian Diagonalization Method (HDM) for obtaining the eigenenergies for the potentials in (1.1) and (1.2). The general Hamiltonian is written as follows

$$\hat{H} = -\frac{1}{2}\frac{d^2}{dx^2} + V(x), \qquad (A1)$$

where $V(x)$ reads

$$V(x) = \begin{cases} \dfrac{V_1}{\cosh^2(\lambda x/2)} + V_2 \cosh(\lambda x), 1D \\ \dfrac{\tilde{V}_0}{\sinh^2(\lambda r/2)} + \dfrac{\tilde{V}_1}{\cosh^2(\lambda r/2)} + \tilde{V}_2 \cosh(\lambda r), 3D \end{cases}, \qquad (A2)$$

Next, we use the same basis set define in Section 2, which read

$$\chi_n(y) = A_n (y-1)^{\frac{2\mu+1}{4}} (y+1)^{\frac{2\nu+1}{4}} P_n^{(\mu,\nu)}(y), \qquad (A3)$$

Now, we define the matrix elements of the Hamiltonian (A1) in this basis as

$$H_{n,m} = \int_{x_-}^{x_+} \chi_n(y) \hat{H} \chi_m(y) dx, \tag{A4}$$

Due to the structure of the 1D and 3D potentials, the Hamiltonian matrix (A4) is split into a sum of tridiagonal and non-tridiagonal matrices. The evaluation of the latter matrix is based on Gauss Quadrature approximation [22-24]. Once the form of the Hamiltonian matrix is obtained, it can be easily diagonalized numerically, and thus the eigenvalues are obtained.

**Figure Captions**

**Figure 1**: Plot of the effective potential in (3.1) for $V_0=1$, $V_1=-50$, $V_2=10$ and $\lambda=1$ for different choices of the angular momentum as indicated.

**Figure 2**: Plot of the potential (1.1) for different choices of the potential parameters showing that the potential could support only bound states as in (a) or it could support only resonances as in (c) or none as in (b, d).

**Figure 3**: Plot of the potential (1.2) for $V_2=10$, $\lambda=1$ and $V_1=-\frac{1}{4}(v^2-1/4)$, where the values of $v$ are indicated in the legends.

**Figure 4**: A plot of the ground, first excited, second excited, and third excited states unnormalized wavefunctions for the 1D hyperbolic well for $V_1=-100$, $V_2=5$, and $\lambda=\sqrt{2}$.

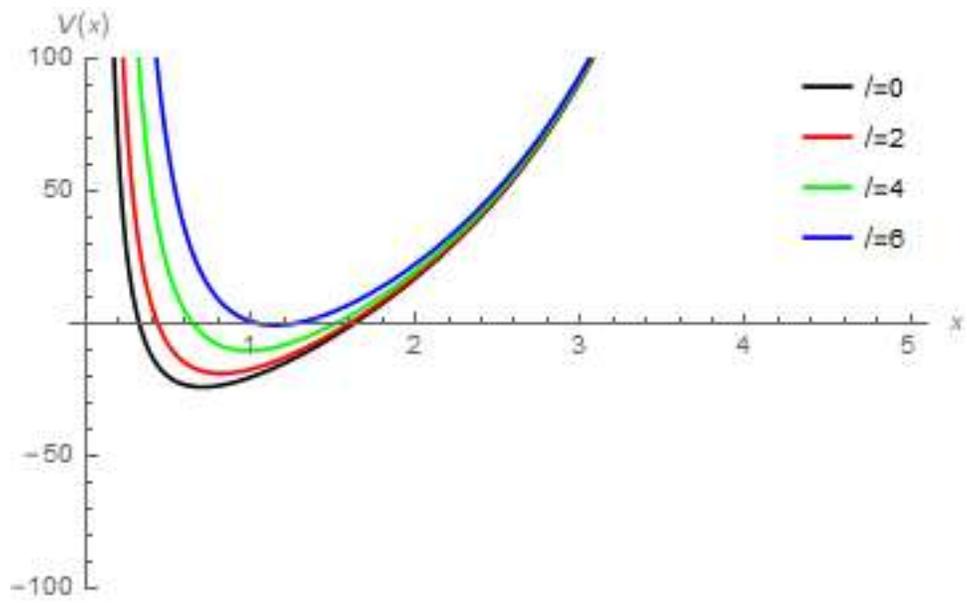

**Figure 1**

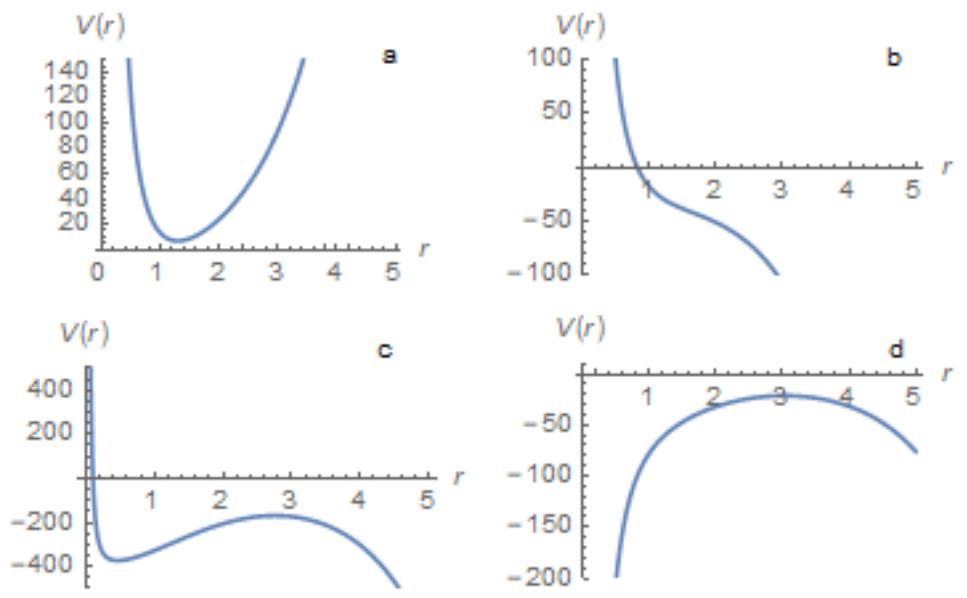

**Figure 2**

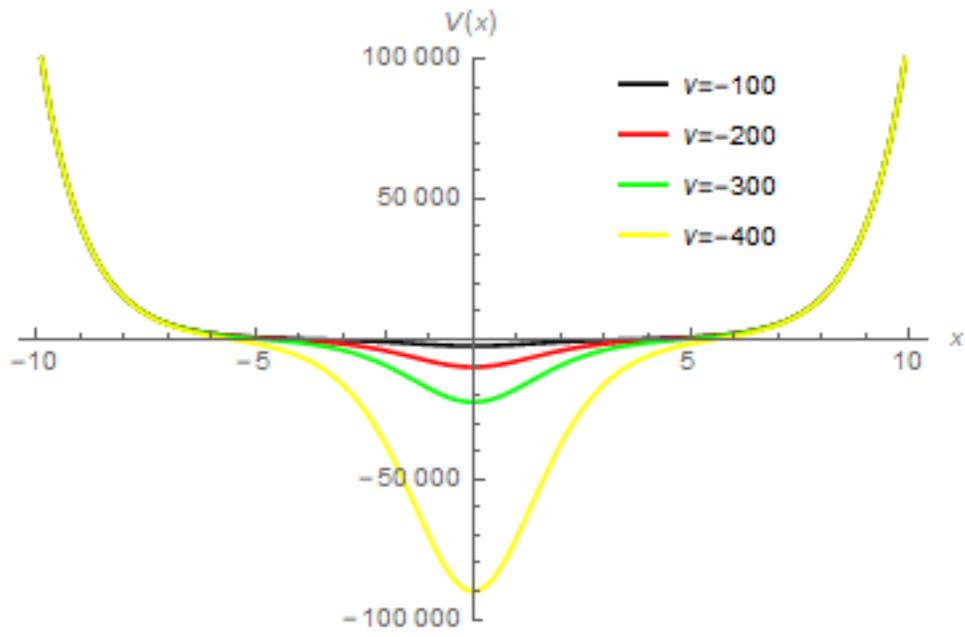

**Figure 3**

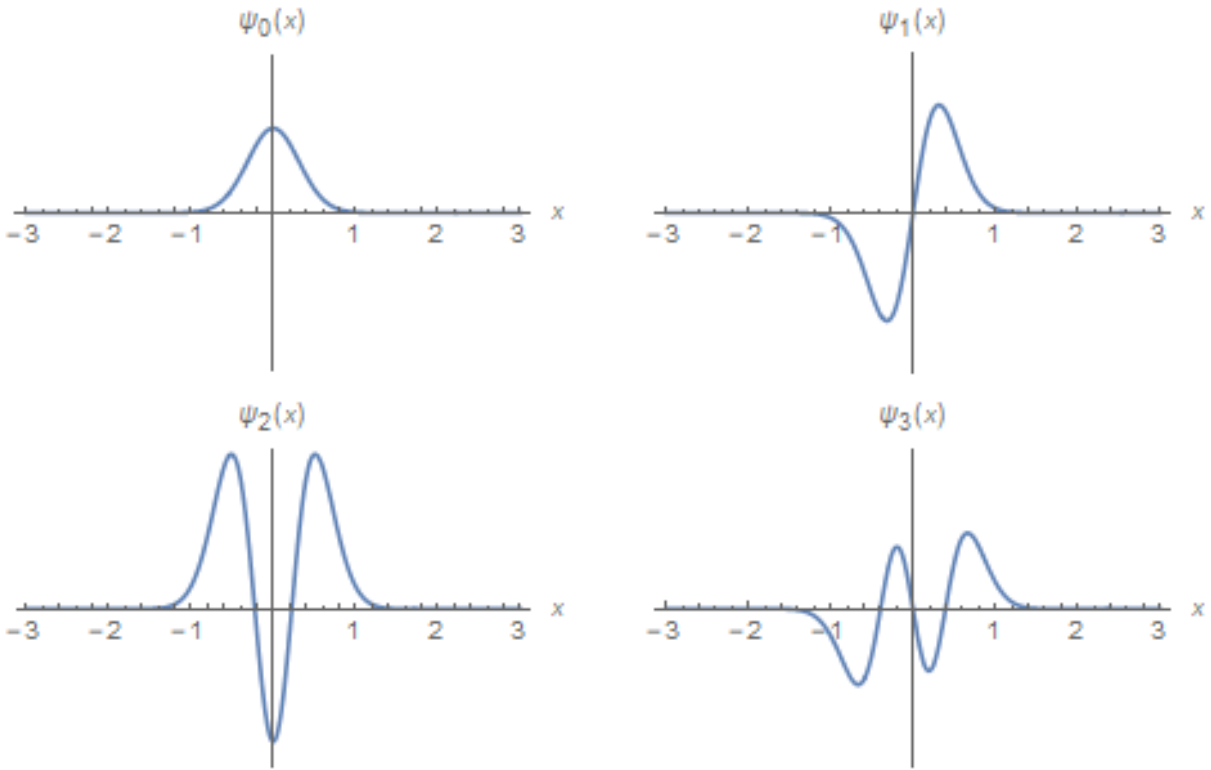

**Figure 4**

**Table Captions**

**Table 1**: The lowest few bound state energies for the potential (1.1) versus the basis size $N=20$, 30, 50, 100. The parameters have been taken as $V_0 = 10$, $V_1 = -200$, $V_2 = 10$, and $\lambda = \sqrt{2}$ for the $s$-wave case, i.e. $\ell = 0$.

**Table 2**: The first 10 eigen energies for the $\ell = 5$ state where the potential parameters have been taken as $V_0 = 10$, $V_1 = -200$, $V_2 = 10$, and $\lambda = \sqrt{2}$.

**Table 3**: The lowest few energies for the first ten even states computed numerically for the 1D hyperbolic potential well (1.2) for $V_1 = -100$, $V_2 = 5$, and $\lambda = \sqrt{2}$ ($\mu = -1/2$).

**Table 4**: The lowest ten odd states energies for the 1D hyperbolic well for $V_1 = -100$, $V_2 = 5$, and $\lambda = \sqrt{2}$ ($\mu = +1/2$).

**Table 5**: The convergence of the lowest few energies computed numerically for the potential (1.2) using Eqs. (3.4) and (3.5) for the potential parameters $V_1 = -10000$, $V_2 = 5$, and $\lambda = \sqrt{2}$.

**Table 1**

| n | $20 \times 20$ | $30 \times 30$ | $40 \times 40$ | $50 \times 50$ |
|---|---|---|---|---|
| 0 | -92.7542191070 | -92.7542191071 | -92.7542191071 | -92.7542191071 |
| 1 | -69.5274752304 | -69.5274752304 | -69.5274752304 | -69.5274752304 |
| 2 | -47.4599081966 | -47.4599081966 | -47.4599081966 | -47.4599081966 |
| 3 | -26.3521411922 | -26.3521411922 | -26.3521411922 | -26.3521411922 |
| 4 | -5.9920899481 | -5.9920899479 | -5.9920899479 | -5.9920899479 |
| 5 | 13.8293062910 | 13.8293062858 | 13.8293062858 | 13.8293062858 |
| 6 | 33.3030737319 | 33.3030735452 | 33.3030735452 | 33.3030735452 |
| 7 | 52.5922960163 | 52.5923160319 | 52.5923160320 | 52.5923160320 |
| 8 | 71.8292716943 | 71.8289092154 | 71.8289091920 | 71.8289091921 |
| 9 | 91.1121245235 | 91.1153393202 | 91.1153398126 | 91.1153398126 |

**Table 2**

| $n$ | $20\times 20$ | $30\times 30$ | $40\times 40$ | $50\times 50$ |
|---|---|---|---|---|
| 0 | -67.4183920149 | -67.4183920149 | -67.4183920149 | -67.4183920149 |
| 1 | -45.5096792463 | -45.5096792463 | -45.5096792463 | -45.5096792463 |
| 2 | -24.5678690265 | -24.5678690265 | -24.5678690265 | -24.5678690265 |
| 3 | -4.3773419049 | -4.3773419049 | -4.3773419048 | -4.3773419048 |
| 4 | 15.2748216788 | 15.2748216809 | 15.2748216809 | 15.2748216809 |
| 5 | 34.5830154742 | 34.5830153204 | 34.5830153204 | 34.5830153204 |
| 6 | 53.7126913183 | 53.7126959922 | 53.7126959924 | 53.7126959924 |
| 7 | 72.7969917087 | 72.7969867948 | 72.7969867879 | 72.7969867879 |
| 8 | 91.9397069063 | 91.9387272192 | 91.9387272103 | 91.9387272103 |
| 9 | 111.1960119413 | 111.2152699336 | 111.2152735975 | 111.2152735964 |

**Table 3**

| $n$ | $15\times 15$ | $20\times 20$ | $30\times 30$ | $50\times 50$ |
|---|---|---|---|---|
| 0 | -89.8612818110 | -89.8612818110 | -89.8612818110 | -89.8612818109 |
| 1 | -70.1359637166 | -70.1359637166 | -70.1359637166 | -70.1359637166 |
| 2 | -51.9825141858 | -51.9825141858 | -51.9825141858 | -51.9825141858 |
| 3 | -35.2237088864 | -35.2237088863 | -35.2237088863 | -35.2237088863 |
| 4 | -19.6260386732 | -19.6260386805 | -19.6260386805 | -19.6260386805 |
| 5 | -4.9111271804 | -4.9111263014 | -4.9111262992 | -4.9111262992 |
| 6 | 9.2122681025 | 9.2122538255 | 9.2122537047 | 9.2122537047 |
| 7 | 23.0088223702 | 23.0085087311 | 23.0085085625 | 23.008508563 |
| 8 | 36.6620554503 | 36.6887073722 | 36.6889029691 | 36.6889029437 |
| 9 | 50.6292842113 | 50.4094241128 | 50.4068618036 | 50.4068617266 |

**Table 4**

| $n$ | $15\times15$ | $20\times20$ | $30\times30$ | $50\times50$ |
|---|---|---|---|---|
| 0 | -79.7931821718 | -79.7931821718 | -79.7931821718 | -79.7931821718 |
| 1 | -60.8725193053 | -60.8725193053 | -60.8725193053 | -60.8725193053 |
| 2 | -43.4420839706 | -43.4420839706 | -43.4420839706 | -43.4420839706 |
| 3 | -27.2963654125 | -27.2963654123 | -27.2963654123 | -27.2963654123 |
| 4 | -12.1766206038 | -12.1766206450 | -12.1766206450 | -12.1766206450 |
| 5 | 2.2069305661 | 2.2069354705 | 2.2069354803 | 2.2069354803 |
| 6 | 16.1368097000 | 16.1366698267 | 16.1366691263 | 16.1366691263 |
| 7 | 29.8506597167 | 29.8523016409 | 29.8523145052 | 29.8523145060 |
| 8 | 43.5020382090 | 43.5352990179 | 43.5356441690 | 43.5356440685 |
| 9 | 57.6577892247 | 57.3207782466 | 57.3142699709 | 57.3142707495 |

**Table 5**

| $n$ | $10\times10$ | $20\times20$ | $30\times30$ | $50\times50$ |
|---|---|---|---|---|
| 0 | -9945.09966135 | -9945.09966135 | -9945.09966135 | -9945.09966135 |
| 1 | -9746.49677011 | -9746.49677011 | -9746.49677011 | -9746.49677011 |
| 2 | -9549.8907267 | -9549.8907267 | -9549.8907267 | -9549.8907267 |
| 3 | -9355.28140035 | -9355.28140035 | -9355.28140035 | -9355.28140035 |
| 4 | -9162.66865346 | -9162.66865346 | -9162.66865346 | -9162.66865346 |
| 5 | -8972.05234112 | -8972.05234112 | -8972.05234112 | -8972.05234112 |
| 6 | -8783.4323107 | -8783.4323107 | -8783.4323107 | -8783.4323107 |
| 7 | -8596.80840134 | -8596.80840134 | -8596.80840134 | -8596.80840134 |
| 8 | -8412.18044335 | -8412.18044336 | -8412.18044336 | -8412.18044336 |
| 9 | -8229.5464644 | -8229.54825775 | -8229.54825775 | -8229.54825775 |